\documentclass[oldversion]{aa}

\usepackage{txfonts}
\usepackage{graphicx}
\usepackage{natbib}
\bibpunct{(}{)}{;}{a}{}{,}

\begin{document}
\defcitealias{P2006002}{Paper~I}
\defcitealias{A2003009}{MMSS00}
\defcitealias{A2007011}{MMSS04}

\title{A comprehensive study of Cepheid variables\\in the Andromeda
galaxy\thanks{Based on observations made with the Isaac Newton Telescope
operated on the island of La Palma by the Isaac Newton Group in the Spanish
Observatorio del Roque de los Muchachos of the Instituto de Astrof\'{\i}sica de
Canarias.}}
\subtitle{Period distribution, blending and distance determination}

\author{F. Vilardell\inst{1} 
        \and C. Jordi\inst{1,3}
        \and I. Ribas\inst{2,3}} 
\institute{Departament d'Astronomia i Meteorologia, Universitat de Barcelona,
c/ Mart\'{\i} i Franqu\`es, 1, 08028 Barcelona, Spain\\
 \email{[francesc.vilardell;carme.jordi]@am.ub.es}
 \and
  Institut de Ci\`encies de l'Espai--CSIC, Campus UAB, Facultat de Ci\`encies,
Torre C5-parell-2a, 08193 Bellaterra, Spain\\
 \email{iribas@ieec.uab.es}
 \and 
  Institut d'Estudis Espacials de Catalunya (IEEC), Edif. Nexus, c/ Gran
Capit\`a, 2-4, 08034 Barcelona, Spain 
}

\date{Received / Accepted}

\abstract{Extragalactic Cepheids are the basic rungs of the cosmic distance
scale. They are excellent standard candles, although their luminosities and
corresponding distance estimates can be affected by the particular properties
of the host galaxy. Therefore, the accurate analysis of the Cepheid population
in other galaxies, and notably in the Andromeda Galaxy (M\,31), is crucial to
obtaining reliable distance determinations. We obtained accurate photometry (in
$B$ and $V$ passbands) of 416 Cepheids in M\,31 over a five year campaign
within a survey aimed at the detection of eclipsing binaries. The resulting
Cepheid sample is the most complete in M\,31 and has almost the same period
distribution as the David Dunlap Observatory sample in the Milky Way. The large
number of epochs ($\sim250$ per filter) has permitted the characterisation of
the pulsation modes of 356 Cepheids, with 281 of them pulsating in the
fundamental mode and 75 in the first overtone. The period-luminosity
relationship of the fundamental mode Cepheids has been studied and a new
approach has been used to estimate the effect of blending. We find that the
blending contribution is as important as the metallicity correction when
computing Cepheid distance determinations to M\,31 ($\sim0.1$ mag). Since large
amplitude Cepheids are less affected by blending, we have used those with an
amplitude $\mathcal{A}_V>0.8$ mag to derive a distance to M\,31 of
$(m-M)_0=24.32\pm0.12$ mag.}


\maketitle

\section{Introduction}

Cepheids are probably the most studied variable stars. Their large amplitudes
and intrinsic luminosities make them easily detectable in most photometric
variability surveys. In addition, their well-known period-luminosity (P-L)
relationship \citep{Leavitt12} has made these variable stars one of the main
cornerstones in deriving extragalactic distances \citep[see][for an historical
review]{A2005021}. The importance of Cepheids for distance determination stands
in contrast with the relative lack of additional information on the specific
characteristics of extragalactic Cepheids and the possible corrections because
of their particular properties (i.e., metallicity). A clear example is the
Andromeda galaxy (M\,31), where the first identification of Cepheids was
already performed by \citet{A2004017}. After the observations of \citet[][and
references therein]{A2004022}, few efforts have been dedicated to further
analyse the Cepheid population in M\,31.

This trend is changing in recent years with the emergence of new observational
capabilities. Several variability surveys have started to study the stellar
content in M\,31 \citep[][and references therein]{A2004003} and other Local
Group galaxies \citep{A2007006, Mochejska01, Udalski01, Pietrzynski04,
A2006023}, obtaining large samples of Cepheids with accurate photometry. The
detailed study of the observed Cepheids has emphasized the importance of an
issue that was usually overlooked in most photometric studies: the effect of
blending. It has been proposed \citep[][hereafter
\citetalias{A2003009}]{A2003009} that the magnitude of Cepheids may be affected
by the light of unresolved companion stars (i.e., blends). The effect of
blending is somewhat different from crowding or confusion noise, since
companion stars appear to be in the same point-like source. Therefore, even
when achieving a perfect point-spread function modeling, blending could still
be present. The effect can be the same as in spectroscopic binaries, where the
individual components cannot usually be resolved from ground-based images.

When studying extragalactic Cepheids the spatial resolution decreases linearly
with the distance of the host galaxy and, as a consequence, the number of
possible blends increases. Small blending values are expected for the Large
Magellanic Cloud, where individual Cepheids can be resolved from neighboring
stars, even from ground-based observations. The situation changes in M\,31 and
M\,33, where mean blending values of $0.18$ mag and $0.16$ mag, respectively,
have been obtained \citep[][hereafter
\citetalias{A2007011}]{A2007011}. These results would imply, when extrapolated
to more distant galaxies, a downward revision of the Hubble constant between
5\% to 10\% \citep[as explained by][]{A2007008}. By contrast, \citet{A2007004}
found an upper limit on blending in NGC\,300 (at $\sim2$ Mpc) of $0.04$ mag. In
addition, \citet{A2007008} showed that the systematic effect of blending on
farther galaxies (between 4 and 25 Mpc) is almost negligible.  Therefore,
results obtained so far seem to indicate that blending has an important
contribution when observing Local Group galaxies and diminishes when observing
distant galaxies. \citet{A2007008} explained such behavior as a consequence of
the background levels in M\,31 and M\,33 (with a large number of stars detected
around Cepheid variables) not being representative of the more distant galaxies
(where the background levels are the result of several unresolved sources).
Because of the importance of the subject, a comprehensive study of the effect
of blending on Cepheid distance determinations is highly desirable.

One of the most recent variability surveys in M\,31 was carried out by us and
was centered on the North-Eastern quadrant of the galaxy \citep[][hereafter
\citetalias{P2006002}]{P2006002}. Our main goal was to detect eclipsing
binaries, although a sample of 416 Cepheids also resulted from the reduction
and analysis process. The large number of detected stars and the high quality
of the resulting light curves have allowed the current study of Cepheid
properties in M\,31.

\section{Photometric sample}
\label{secphot}

\begin{figure}[!tb]
 \resizebox*{\hsize}{!}{\includegraphics{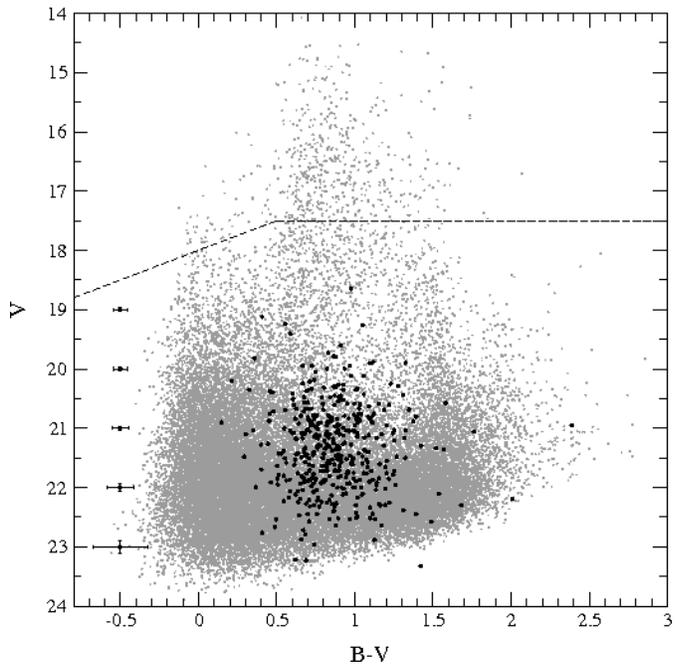}}
 \caption{Color-magnitude diagram for 37\,241 objects with photometric errors
below 0.1 mag in the reference catalog of \citetalias{P2006002} (grey dots) and
the 416 identified Cepheids (black circles). Objects above the dashed line are
saturated. Error bars on the left indicate the mean error, in magnitude and
color, at different magnitude values.}
 \label{vbv}
\end{figure}

The photometric survey in \citetalias{P2006002} was carried out with the 2.5~m
Isaac Newton Telescope (INT) at La Palma (Spain) over the course of five
observing runs (between 1999 and 2003). An implementation of the difference
image analysis technique \citep{A2002001} was used in the reduction process to
automatically detect 3\,964 variable stars with $\sim$250 observations per band
($B$ and $V$). The analysis of variance algorithm \citep{A2003002} was used to
identify periodic variable stars. The folded light curves were visually
inspected and lead to the identification of 437 eclipsing binaries and 416
Cepheids. The photometric measurements were transformed to the standard system
through the observation of several \citet{A2004027} star fields, obtaining
magnitude errors between 0.03 and 0.11 mag. Finally, the phase-weighted mean
magnitudes \citep{A2005001} of the Cepheid sample were determined for each
passband ($B$ and $V$). Figure \ref{vbv} shows the color-magnitude diagram of
the observed stars with good photometric precision. Cepheid variables are
highlighted.

\section{Period distribution}
\label{secper}

The period distribution of Cepheids in Local Group galaxies has been a major
source of debate. Evolution models predict a displacement on the peak of the
period distribution as a function of the metallicity of the host galaxy
\citep{A2006001}. It has also been observed that the Milky Way (MW) period
distribution displays a dip at around 10 days, while such feature is missing in
more metal-poor galaxies such as the LMC. Several theories have been put
forward to explain the bimodal distribution, but a satisfying scenario is still
lacking \citep[see][for an extended discussion]{A2004020}.

Since the MW and M\,31 have similar morphological type and chemical
composition, the observed period distributions are expected to be also similar.
The results obtained so far have prevented a direct comparison because of
observational biases, which are often difficult to evaluate. On the one hand,
faint Cepheids (with usually short periods) can be missing in some shallow
surveys and, on the other hand, the distribution of the observations can make
the identification of long period Cepheids difficult.

\begin{figure}[!tb]
 \resizebox{\hsize}{!}{\includegraphics*{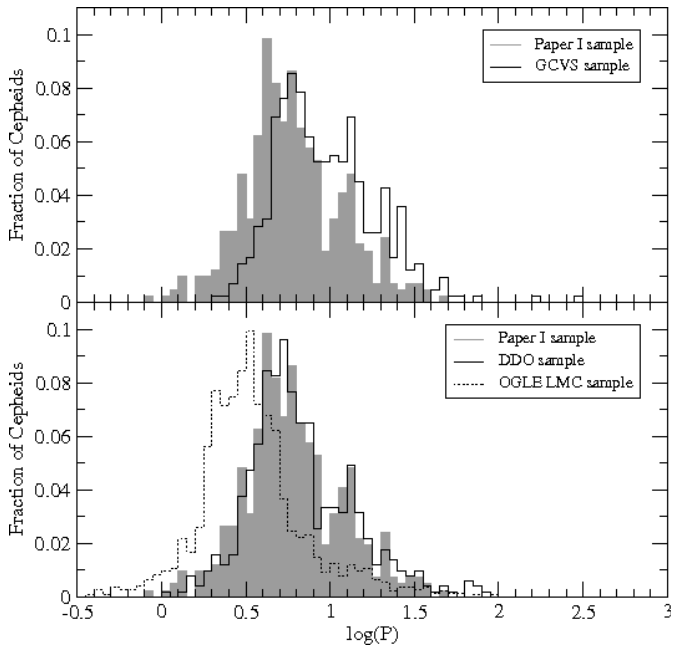}}
 \caption{\emph{Top}: Normalized period distribution for the 416 detected 
Cepheids in \citetalias{P2006002} compared with the 420 Cepheids in M\,31 
of the GCVS. \emph{Bottom}: Normalized period distribution for the 416 
detected Cepheids in M\,31 compared with the 509 Cepheids in the MW from 
the David Dunlap Observatory (DDO) sample. The OGLE Large Magellanic Cloud 
Cepheid sample is also shown for comparison.}
 \label{pdist}
\end{figure}

To evaluate the observational biases, the 416 Cepheids from
\citetalias{P2006002} were compared with the 420 Cepheids in M\,31 of the GCVS
\citep{Samus04}. Both samples have almost the same number of stars, but the two
period distributions (Fig. \ref{pdist}) are found to be largely different, as
demonstrated by the Kolmogorov-Smirnov tests, which provide values of around
$10^{-13}$. The origin of the observed difference can be explained by two main
factors. Firstly, the \citetalias{P2006002} sample is known to have a bias for
the longest period Cepheids because of the observational window function (see
\citetalias{P2006002} for further information). And secondly, the GCVS presents
an important bias at short periods because the GCVS is shallower than our
survey and faint Cepheids are missing. Furthermore, both Cepheid samples in
M\,31 can be compared with the MW sample of the David Dunlap
Observatory\footnote{Data obtained from:
\texttt{http://www.astro.utoronto.ca/DDO/\\research/cepheids/cepheids.html}}
\citep[DDO,][]{A2007002}. As can be seen (Fig. \ref{pdist}), the period
distributions of \citetalias{P2006002} and the DDO sample are very similar
(with a Kolmogorov-Smirnov test value of 0.49). Only a slight difference is
observed at long periods, possibly because of the observational bias in the
\citetalias{P2006002} sample. Therefore, as expected, equivalent period
distributions are obtained for similar galaxies, including the secondary peak
at $P>10$ d in metal rich galaxies.

The conspicuous similarity between the period distributions of MW and M\,31
becomes even more striking when compared with a galaxy of different metallicity
(Fig. \ref{pdist}). More than 1300 Cepheids have been observed in LMC as part
of the OGLE II survey\footnote{Data obtained from:
\texttt{ftp://sirius.astrouw.edu.pl/ogle/\\ ogle2/var\_stars/lmc/cep/catalog/}}
\citep{A2007006}. The resulting period distribution reveals a single peak
which, as expected, is shifted towards shorter periods with respect to those of
the distributions of the MW and M\,31. Therefore, the great similarity between
the DDO and \citetalias{P2006002} samples indicates that both the observational
biases and period distributions are equivalent.

\section{Fourier decomposition}
\label{secfou}

It is well known that Fourier decomposition of the Cepheid light curves can
provide valuable information on the real nature of these variable stars,
especially on their pulsation modes \citep{A2005032}. The procedure involves
fitting the coefficients ($A_k$ and $\varphi_k$) of a Fourier series of the
form:
\begin{equation}
 \label{eqfourier}
 m(t)=A_0+\sum^{J}_{k=1}A_k\cos\left(2\pi k\Phi(t)+\varphi_k\right)
\end{equation}
where $m(t)$ and $\Phi(t)$ are the magnitude of the Cepheid and the phase at
time $t$, respectively. Once the best coefficients have been obtained, the
pulsation mode of the Cepheid variable can be derived from the amplitude ratio
$R_{k1}=A_k/A_1$ and phase difference $\varphi_{k1}=\varphi_k-k\varphi_1$.

A prerequisite of the Fourier decomposition of the 416 Cepheids identified in
M\,31 is the determination of the maximum order of the fit, $J$.  Similarly to
other studies in the literature \citep{A2005029, A2005032, A2005030}, we
observed that an iterative solution provided satisfactory results. Therefore,
for each light curve, the fitting procedure started with $J=1$ and increased
until the value of $\chi^2_{dof}$ varied by less than one. Although the Fourier
fit always provided values for all the light curves, several important aspects
were considered before studying the results. Firstly, since measurements with
large photometric errors can provide unreliable fits, light curves with a mean
photometric error larger than 0.1 mag were excluded. Secondly, all light curves
with $\chi^2_{dof}>7$ were rejected to eliminate the fits that did not
accurately match the observations. Finally, large gaps in a light curve can
prevent the fit to correctly reproduce its shape, even when $\chi^2_{dof}$ is
relatively low. Problematic fits usually have coefficients with large error
bars. Therefore, coefficient errors were computed \citep[according
to][]{A2006027} and light curves with $\sigma_{A_0}>0.01$ mag,
$\sigma_{R_{21}}>0.1$ or $\sigma_{\varphi_{21}}>0.6$ rad ($\simeq0.2\pi$ rad)
were also rejected.

\begin{figure}[!tb]
 \resizebox{\hsize}{!}{\includegraphics*{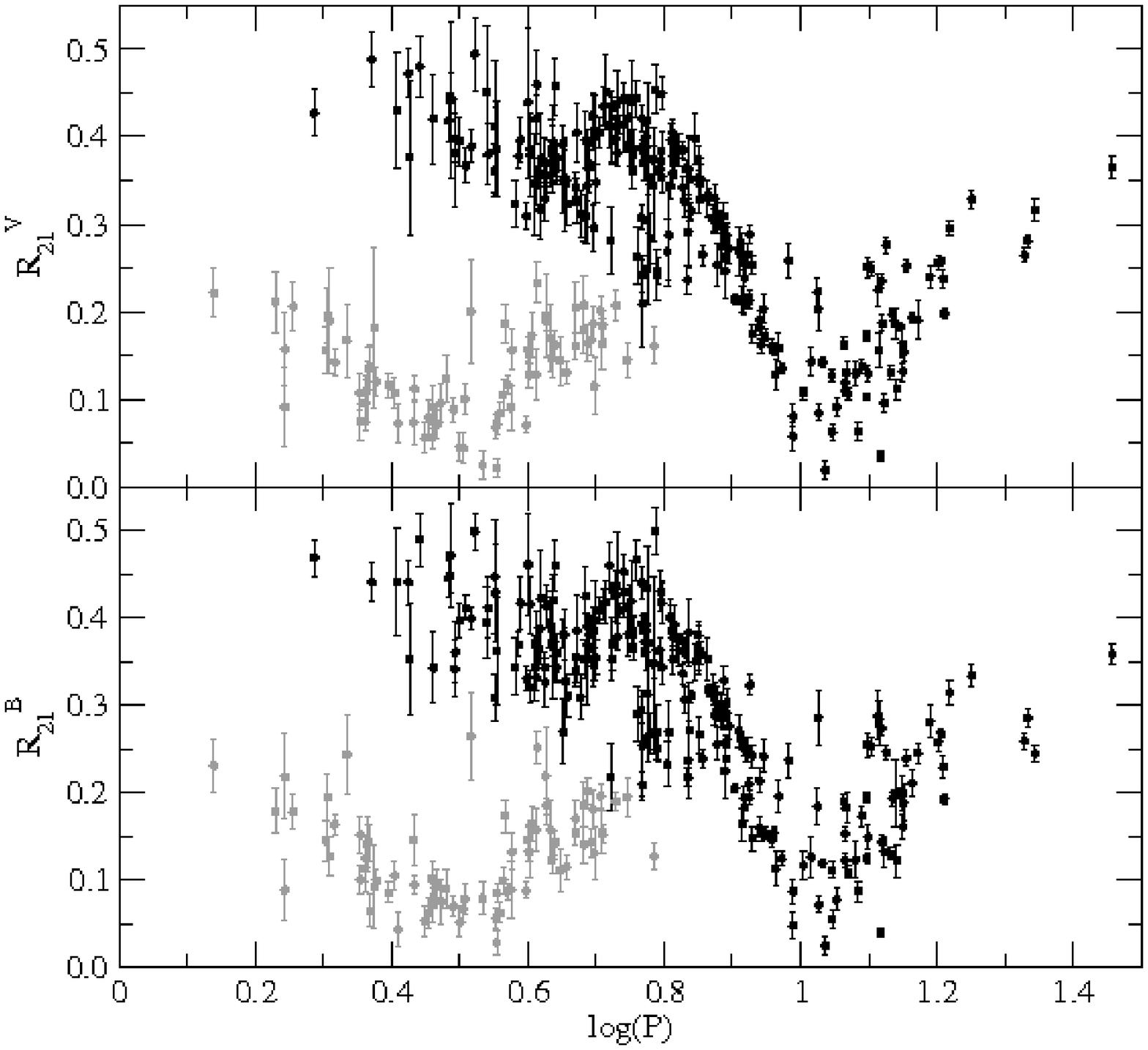}} 
 \resizebox{\hsize}{!}{\includegraphics*{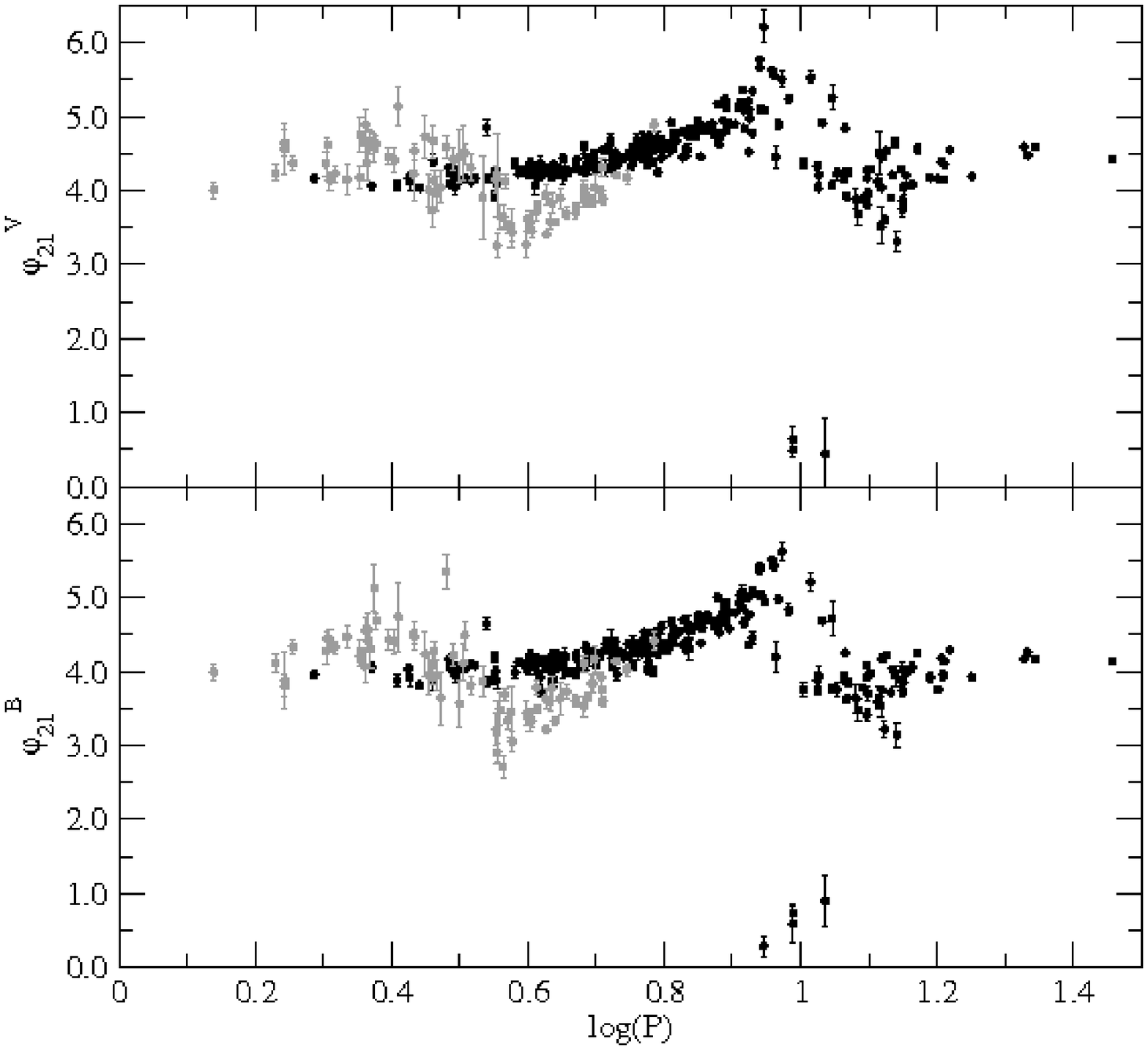}}
 \caption{Fourier coefficients in each passband ($B$ and $V$) as a 
function of period. The pulsation mode of the 315 Cepheids is identified: 
black circles -- fundamental mode; grey circles -- first overtone.}
 \label{fourier}
\end{figure}

The above criteria yielded a sample of 315 Cepheids with accurate Fourier fits
in both passbands ($B$ and $V$). The $R_{21}$ and $\varphi_{21}$ values in
\emph{both passbands} (Fig. \ref{fourier}) have been used to classify 75
Cepheids pulsating in the first-overtone mode (FO) and 240 pulsating in the
fundamental mode (FM). Since FO Cepheids are not expected to have periods
longer than $\sim 7$ days, the FM sample has been completed with all Cepheids
with $\log P>0.9$ ($\simeq 8$ days), regardless of the quality of their Fourier
fit. The final sample of 281 FM Cepheids includes three type II Cepheid
candidates, but they are easily identified as a consequence of the detailed
analysis of the P-L distribution (Sect. \ref{secabs}).

\section{Period-luminosity relationship}
\label{seceff}

\begin{figure}[!tb]
 \resizebox{\hsize}{!}{\includegraphics*{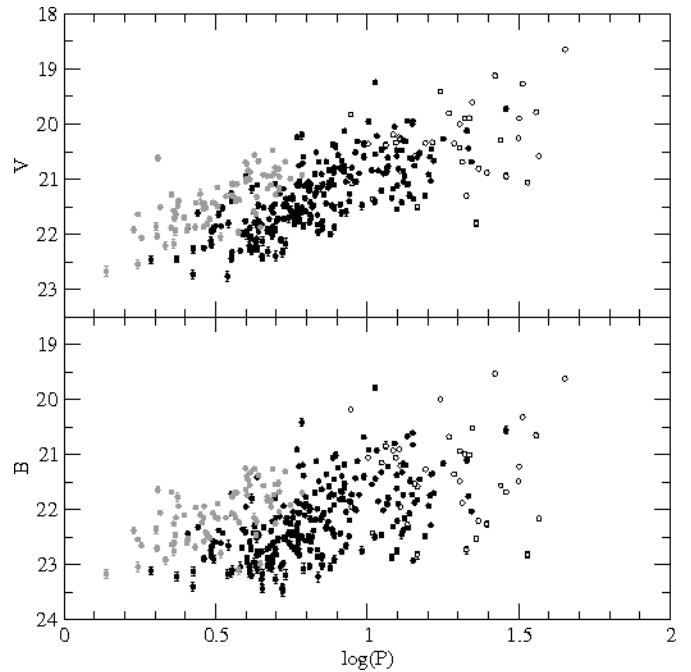}}
 \caption{Observed magnitudes as a function of period for 356 Cepheids. Filled
black circles: Fundamental mode Cepheids with accurate Fourier coefficients.
Empty black circles: Fundamental mode Cepheids without accurate Fourier
coefficients. Grey circles: First overtone mode pulsators.}
 \label{plum}
\end{figure}

The P-L diagram for the 356 classified Cepheids (281 FM and 75 FO) reveals a
large scatter in both passbands (Fig. \ref{plum}), especially in $B$.  The
origin of the observed scatter can be understood if we consider that the field
of view covers about 7 kpc at the distance of M\,31
\citep[$(m-M)_0=24.44\pm0.12$ mag,][]{P2005002}. The large covered area in
M\,31, the fact that Cepheids are located along the spiral arms, and the clumpy
structure of the interstellar medium (clearly observed in the survey images),
indicates that differential absorption in the disk is the most likely
responsible for an important fraction of the observed scatter. In addition, the
larger dispersion of the P-L distribution in $B$ compared to $V$ reinforces the
interstellar absorption hypothesis.

Another possible effect on the observed scatter is metallicity. Several authors
have suggested \citep[see, i.e.,][]{A2007016} that the slope of the Cepheid P-L
relationship is basically independent of metallicity.  However, comparison of
tip of the red giant branch and Cepheid distance determinations to several
galaxies has shown that a slight dependence on metallicity is present
\citep{A2007014}. Therefore, a metallicity dependence on the zero point of the
Cepheid P-L relationship could exist.  Since metallicity within M\,31 decreases
as a function of the galactocentric distance \citep{A2007012}, some of the
observed scatter could be as well introduced by metallicity differences.

Also, the Cepheid measured photometry can be affected by blends. The mean
seeing of our images is around 1~arcsec. Therefore, each observed point-like
source corresponds to $\sim 4$ pc at the distance of M\,31.  Since Cepheids are
usually located in young star clusters and associations, it is likely that the
observed magnitude of a Cepheid can be the combination of several unresolved
sources.

Below we analyse the three aforementioned effects that are the probable sources
of the scatter in the P-L diagram.

\subsection{Absorption}
\label{secabs}

The effect of differential absorption can be partially corrected from the
observed Cepheid color ($B-V$) through the color-excess
\citep[$E(B-V)\equiv(B-V)-(B-V)_0$,][]{A2003010}. We used P-L relationships
from \citet{A2007006}\footnote{Updated relationships were obtained from the
OGLE II web site:
\texttt{ftp://sirius.astrouw.edu.pl/ogle/ogle2/var\_stars/lmc/\\cep/catalog/README.PL}}
to estimate $B_0$ and $V_0$ values. The LMC distance modulus was assumed to be
$(m-M)_0=18.42\pm0.06$ mag and was computed from the weighted mean of all the
LMC distance determinations with eclipsing binaries \citep[][]{A2003004}. The
reason for using LMC relationships (instead of those for the MW), was motivated
by several recent results. First, \citet{A2007016} suggested that the MW
Cepheid distances, mostly obtained through the Baade-Wesselink method, could be
affected by a systematic bias when converting the observed radial velocities
into pulsation velocities. Second, \citet{A2006023} found that LMC P-L slopes
provide a better agreement with the P-L distribution of Cepheids in NGC\,4258
than MW relationships. Finally, the new parallax measurements of several MW
Cepheids \citep{Benedict07,A2007013} yield P-L slopes that seem to be in better
agreement with the LMC than with previous MW relationships.

\begin{figure}[!tb]
 \resizebox{\hsize}{!}{\includegraphics*{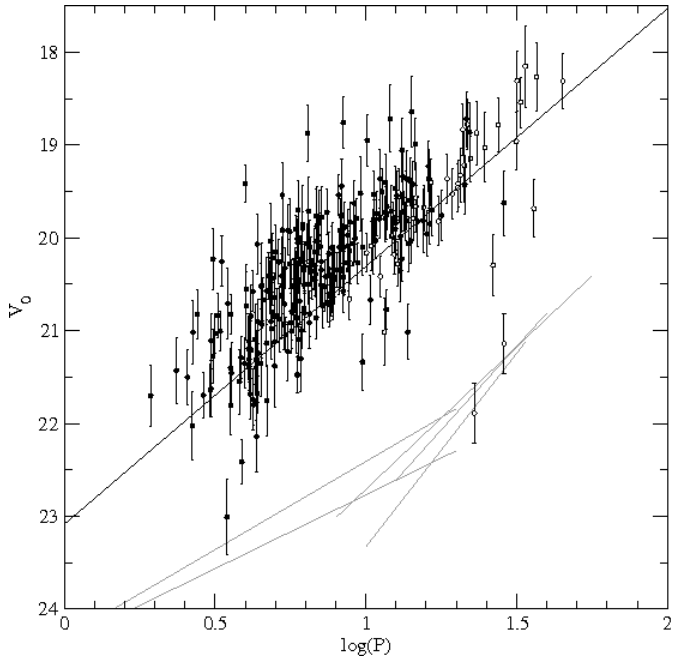}}
 \caption{Absorption-corrected $V$ magnitude as a function of period for 
281 fundamental mode Cepheids. Filled circles: Stars with accurate Fourier 
fit. Empty circles: Stars without accurate Fourier fits. Black line: 
\citet{A2007006} P-L relationship, for a distance modulus to M\,31 of 
24.44 mag \citep{P2005002}. Grey lines: Distance corrected P-L 
relationships for type II Cepheids as given by \citet{Alcock98}.}
 \label{pvo} 
\end{figure}

After obtaining the color excess for each Cepheid, a total-to-selective
extinction ratio of $\mathcal{R}_V\equiv A(V)/E(B-V)=3.1\pm0.3$
\citep{A2005027} was used to compute the absorption and the $V_0$ magnitude for
every Cepheid. The resulting P-L distribution can be compared with the LMC P-L
relationship (Fig. \ref{pvo}), assuming a M\,31 distance modulus of
$(m-M)_0=24.44\pm0.12$ mag \citep{P2005002}. An offset between the $V_0$ values
and the LMC P-L relationship is clearly observed, but the effects of
metallicity and blending have still to be considered.

Finally, as previously mentioned, three of the studied Cepheids seem to be of
type II. To further investigate these objects they have been kept in the
studied sample of FM Cepheids (Sects. \ref{secblend} and \ref{secdist}).

\subsection{Metallicity}
\label{secmetal}

As explained above, the zero point of the P-L relationship may depend on
metallicity. To account for this effect, the metallicity of each Cepheid needs
to be estimated, and we did so by considering a galactocentric metallicity
dependence. According to \citet{A2007012}, the M\,31 galactocentric metallicity
dependence can be modeled by:
\[
12+\log(O/H)=(9.03\pm0.09)-(0.28\pm0.10)(\rho/\rho_0-0.4)
\]
\noindent where $\rho$ is the de-projected galactocentric radius and $\rho_0$
is the isophotal radius ($77.44$ arcmin for M\,31). Following the procedure in
\citet{Baade64}, the Cepheid galactocentric radius was obtained assuming a
position angle of $38\degr$, an inclination of $12\fdg5$ \citep{Simien78} and
an M\,31 center position of $\alpha=00^h42^m44\fs31$
$\delta=+41\degr16\arcmin09\farcs4$ \citep[in J2000.0
coordinates,][]{Cotton99}.

The metallicity dependence of the Cepheid P-L relationship and the
galactocentric metallicity variation implies that a different P-L relationship
must be used for each individual Cepheid when computing the value $V_0-M_V$.
Alternatively, one can compute $V_0-M_V$ by using a universal P-L relationship
and then applying a correction for each Cepheid \emph{a posteriori}.  According
to \citet{A2007014}, the correction on the computed $V_0-M_V$ value is
$\delta(m-M)_0/\delta(O/H)=-0.25\pm0.09$ mag/dex when using the
\citet{A2007006} P-L relationships. Therefore, we computed metallicity
corrections by comparing the assumed metallicity of each M\,31 Cepheid with the
metallicity of the LMC \citep[$12+\log(O/H)=8.5$ dex,][]{A2007014}. The
resulting corrections for the studied sample lie between $-0.15$ mag and
$-0.05$ mag.

\subsection{Blending}
\label{secblend}

The best approach to study the effect of blending given our available data sets
is by using the observed Cepheid amplitudes \citep{A2007003}. The intrinsic
amplitude $\mathcal{A}^i$ of a Cepheid is given by the expression:
\begin{equation}
 \mathcal{A}^i=-2.5\log\left(\frac{f_c}{F_c}\right)
\end{equation}
\noindent where $f_c$ and $F_c$ are the fluxes at minimum and maximum light,
respectively. When blending is present, the observed amplitude $\mathcal{A}$
can be expressed in the following form:
\begin{equation}
 \mathcal{A}=-2.5\log\left(\frac{f_c+f}{F_c+f}\right)
\end{equation}
\noindent where $f$ is the sum of fluxes of all blending sources. From these
two equations, one can write:
\begin{equation}
 \label{blendamp}
 \mathcal{A}=2.5\log\left(\frac{10^{0.4\mathcal{A}^i}+\frac{f}{f_c}}{1+\frac{f}{f_c}}\right)
\end{equation}
Since $\mathcal{A}^i$ and $f/f_c$ are always positive quantities, the observed
amplitude of a blended Cepheid is always smaller than its intrinsic amplitude.
The observed amplitude and the blended mean magnitude \mbox{$<m>$} can be
expressed as:
\begin{eqnarray}
 \label{insimeq}
 \mathcal{A}=m-M \\
 <m>=\frac{m+M}{2}
\end{eqnarray}
\noindent where $m$ and $M$ are the observed magnitudes at minimum and maximum
light of a Cepheid, respectively. In the same way, the intrinsic amplitude and
the blending-free mean magnitude $<m^i>$ can be expressed as:
\begin{eqnarray}
 \mathcal{A}^i=m^i-M^i\\
 <m^i>=\frac{m^i+M^i}{2} 
 \label{endsimeq}
\end{eqnarray}

From equations (\ref{insimeq})-(\ref{endsimeq}) the difference on the mean
magnitude of a given Cepheid as a consequence of blending can be expressed as:
\begin{equation}
 \label{deltam}
 \Delta = <m>-<m^i>=(m-m^i)-\frac{\mathcal{A}-\mathcal{A}^i}{2}
\end{equation}
Considering that $(m-m^i)$ can be defined as:
\begin{equation}
 m-m^i=-2.5\log\left(\frac{f_c+f}{f_c}\right) = -2.5\log\left(1+\frac{f}{f_c}\right)
\end{equation}
\noindent and isolating $f/f_c$ from (\ref{blendamp}), the difference on the
mean magnitude (\ref{deltam}) can be expressed as:
\begin{equation}
 \label{deltamp}
 \Delta = 2.5\log\left(\frac{10^{0.2\mathcal{A}}-10^{-0.2\mathcal{A}}}{10^{0.2\mathcal{A}^i}-10^{-0.2\mathcal{A}^i}}\right)
\end{equation}

Therefore, the variation on the mean magnitude of a given Cepheid because of
blending can be computed from the amplitude of the Cepheid. Although
$\mathcal{A}^i$ and $\Delta$ are unknown quantities, the above equation can be
solved in combination with period-color and P-L relationships. The results
shown below are based on the \citet{A2007006} relationships, but almost
identical results were obtained with other LMC relationships \citep{A2005011}.

The next step of the process involves recalling that
\mbox{$<m>\simeq\overline{m}$}, where $\overline{m}$ is the phase-weighted
intensity-average mean magnitude of the observed Cepheids. Therefore, defining
the blending-free color excess as:
\begin{equation}
 E(B^i-V^i)\equiv(B^i-V^i)-(B-V)_0
\end{equation}
\noindent the blended color excess can be expressed as:
\begin{equation}
 \label{blendexcess}
 E(B-V)=E(B^i-V^i)+\Delta_B-\Delta_V
\end{equation}
\noindent where $\Delta_B=B-B^i$ and $\Delta_V=V-V^i$. Analogously, the
distance modulus can be expressed as:
\begin{equation}
 \label{distdelta}
 (V_0-M_V)=(m-M)_0+\Delta_V-\mathcal{R}_V(\Delta_B-\Delta_V)
\end{equation}
\noindent where: 
\begin{eqnarray}
(V_0-M_V)=V-\mathcal{R}_VE(B-V)-M_V\\
(m-M)_0=V^i-\mathcal{R}_VE(B^i-V^i)-M_V
\end{eqnarray}

The combination of Eqs. (\ref{distdelta}) and (\ref{deltamp}) reveals that the
value of $(V_0-M_V)$ is only a function of the amplitude of the Cepheids and
the distance modulus to M\,31. In addition, the fact that the color excess is
obtained from the observed color of the Cepheids (which is affected by
blending) introduces the color term $(\Delta_B-\Delta_V)$ in Eq.
(\ref{distdelta}). Therefore, and contrary to the intuitive interpretation,
when reddening is computed from the observed color of Cepheids $(B-V)$ the
blended distance modulus can be either larger or smaller than the intrinsic
distance modulus $(m-M)_0$, depending on the values of $(\Delta_B-\Delta_V)$.

\begin{figure}[!tb]
 \resizebox{\hsize}{!}{\includegraphics*{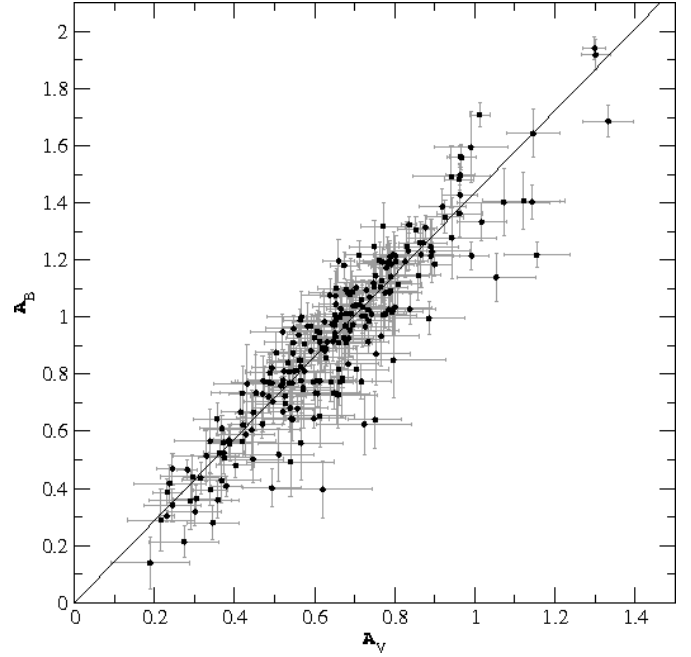}}
 \caption{Amplitude relationship for the 240 fundamental mode Cepheids.}
 \label{avab}
\end{figure}

Considering that all Cepheids are roughly at the same distance, 
$(m-M)_0=24.44\pm0.12$ mag \citep{P2005002}, and assuming that
$\mathcal{A}^i_V$ and $\mathcal{A}^i_B$ are linearly dependent (Fig.
\ref{avab}), Eq. (\ref{distdelta}) can be numerically solved. In fact, when the
intrinsic amplitude of the Cepheid tends to zero, the observed amplitudes have
to be zero in all passbands. Therefore, both amplitudes were considered to be
proportional (i.e., $\mathcal{A}^i_B=\alpha\mathcal{A}^i_V$). The assumption
that Cepheid amplitudes, when observed at different passbands, are proportional
can be considered a first-order approximation of the Fourier coefficient
interrelations \citep{A2005026}.

Since the interrelations between $B$ and $V$ have not been accurately worked
out, we computed the proportionality factor empirically from the observed
amplitude of Cepheids. The amplitude of the studied Cepheids can be reliably
estimated from the Fourier fits (Sect. \ref{secfou}).  Therefore, the 240 FM
Cepheids with accurate Fourier fits were used to compute the amplitude
proportionality factor (Fig. \ref{avab}), obtaining a mean value (with
$2.5\sigma$ clipping) of
$\mathcal{A}^i_B/\mathcal{A}^i_V\simeq\mathcal{A}_B/\mathcal{A}_V=1.435\pm0.011$.

\begin{figure}[!tb]
 \resizebox{\hsize}{!}{\includegraphics*{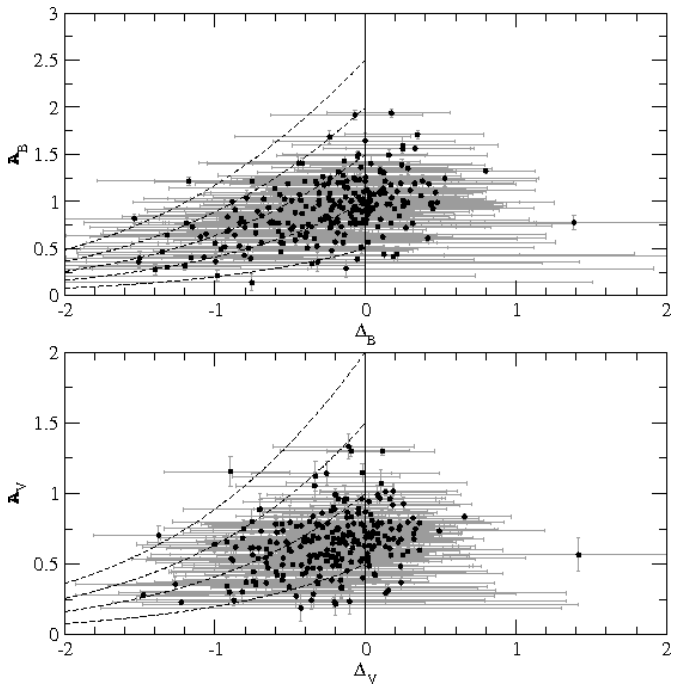}}
 \caption{Observed amplitude of the 240 fundamental mode Cepheids with 
accurate Fourier fits as a function of the computed variation on the mean 
magnitude due to blending. Lines of constant intrinsic amplitude from 0.5 to 
2 or 2.5 mag in steps of 0.5 mag are also shown (dashed lines).}
 \label{deltaamps}
\end{figure}

The above distance modulus to M\,31 and the amplitude proportionality factor,
when applied to Eq. (\ref{distdelta}), yielded the intrinsic amplitudes of the
Cepheids. Once the intrinsic amplitudes are known, the difference on the mean
magnitude can be computed (Fig. \ref{deltaamps}). The large resulting
uncertainties in $\Delta_B$ and $\Delta_V$ clearly indicate that not much can
be said from the individual blending of each Cepheid. The uncertainties were
obtained from a MonteCarlo run with 1000 realisations on the input parameters
of each Cepheid. The positive $\Delta$ values (implying negative blending)
provide additional information on the associated uncertainties (probably of the
order of 0.1-0.2 mag). In any case, it is interesting to observe that moderate
intrinsic amplitudes are obtained for the entire sample, as it can be deduced
from the lines of constant intrinsic amplitude in Fig. \ref{deltaamps} (at
$\Delta=0$, $\mathcal{A}=\mathcal{A}^i$). Furthermore, the Cepheid with
$\Delta_V>1$ mag is the only suspected type II Cepheid with an accurate Fourier
fit (see Fig. \ref{pvo}). The unrealistic blending value provides an additional
evidence in favor of the type II classification.

Given the large number of studied Cepheids, the obtained blending values can
also be used to determine a mean blending value for Cepheids in M\,31.  The
results are shown in Table \ref{blends} where they are also compared with
previous blending determinations in M\,31 and M\,33. The values in
\citetalias{P2006002} were obtained from the third light contribution in
eclipsing binary systems, and the \citetalias{A2003009} and
\citetalias{A2007011} values were obtained from comparison of Hubble Space
Telescope (HST) and ground based images. The blending factors ($S$)
presented in \citetalias{A2003009} and \citetalias{A2007011} were
transformed into variation on the mean magnitude values by considering that:
\begin{equation}
 \label{deltas}
 \Delta = -2.5\log\left(1+\frac{f}{<f_c>}\right) \equiv -2.5\log\left(1+S\right)
\end{equation}
\noindent where $<f_c>$ is the flux of the Cepheid at mean magnitude. 

\begin{table*}[!htb]
 \caption{Blending values in M\,31 and M\,33 obtained from several
methods. \emph{Top:} Mean values and their errors. \emph{Bottom:} Median
values.}
 \label{blends}
 \begin{tabular}{llccccc}
  \hline
  \hline
  Reference & Sample & Number & $<S_V>$ & $<S_B>$ & $<\Delta_V>$ & $<\Delta_B>$ \\
  & & & & & [mag] & [mag] \\ 
  \hline
  \citetalias{A2003009} & Cepheids in M\,31 & 22 (10 in $B$) & $0.19\pm0.03$ & $0.06\pm0.06$ & $-0.18\pm 0.03$ & $-0.05\pm0.05$ \\
  \citetalias{P2006002} & Eclipsing binaries & 48 & $0.31\pm0.07$ & $0.30\pm0.06$ & $-0.23\pm0.05$ & $-0.25\pm0.04$ \\
  Present work & FM Cepheids & 240 & $0.31\pm0.03$ & $0.37\pm0.04$ & $-0.23\pm0.02$ & $-0.24\pm0.03$ \\
  Present work & FM Cepheids with $P > 12$ days & 37 & $0.14\pm0.06$ & $0.20\pm0.10$ & $-0.09\pm0.05$ & $-0.10\pm0.07$ \\
  Present work & FM Cepheids with $\mathcal{A}_V>0.8$ mag & 37 & $0.10\pm0.05$ & $0.11\pm0.07$ & $-0.06\pm0.04$ & $-0.05\pm0.06$ \\
  & & & & & & \\
  \citetalias{A2007011} & Cepheids in M\,33 & 95 (57 in $B$) & $0.24\pm0.03$ & $0.29\pm0.06$ & $-0.20\pm0.02$ & $-0.23\pm0.04$\\
  \citetalias{A2007011} & Cepheids in M\,33 with $P > 10$ days & 60 (39 in $B$) & $0.16\pm0.04$ & $0.20\pm0.05$ & $-0.14\pm0.025$ & $-0.17\pm0.04$ \\

  \hline
  Reference & Sample & Number & Median$(S_V)$ & Median$(S_B)$ & Median$(\Delta_V)$ & Median$(\Delta_B)$ \\ 
  & & & & & [mag] & [mag] \\ 
  \hline
  \citetalias{A2003009} & Cepheids & 22 (10 in $B$) & $0.12$ & $0.00$ & $-0.12$ & $0.00$ \\
  \citetalias{P2006002} & Eclipsing binaries & 48 & $0.09$ & $0.16$ & $-0.09$ & $-0.16$ \\
  Present work & FM Cepheids & 240 & $0.20$ & $0.15$ & $-0.19$ & $-0.16$ \\
  Present work & FM Cepheids with $P > 12$ days & 37 & $0.09$ & $0.04$ & $-0.09$ & $-0.05$ \\
  Present work & FM Cepheids with $\mathcal{A}_V>0.8$ mag & 37 & $0.09$ & $0.04$ & $-0.09$ & $-0.05$ \\
  & & & & & & \\
  \citetalias{A2007011} & Cepheids in M\,33 & 95 (57 in $B$) & $0.13$ & $0.15$ & $-0.13$ & $-0.15$ \\
  \citetalias{A2007011} & Cepheids in M\,33 with $P > 10$ days & 60 (39 in $B$) & $0.07$ & $0.10$ & $-0.07$ & $-0.10$ \\
  \hline
 \end{tabular}
\end{table*}

The observed difference with \citetalias{A2003009} could be due to the assumed
distance moduli or to the observing conditions. On the one hand, the reported
uncertainties on the LMC and M\,31 distance moduli could produce a variation of
$0.12$ mag in $\Delta_B$ and $\Delta_V$. On the other hand, blending depends on
seeing conditions and background level. Large seeing images increase the
blending and high signal-to-noise data makes that faint stars are detected,
decreasing the computed background and increasing the blending contribution.
Considering that similar results have been obtained for the eclipsing binary
sample (which comes from the same observational data but from a completely
different procedure), the obtained differences with the values reported in
\citetalias{A2003009} could be the result of different observing conditions.
The method used by \citetalias{A2003009} can only provide lower limits to
blending values, since the HST point spread function (used as
blending-free reference) could still hide unresolved companions.
This is especially the case for the $B$-band results, where the 
available HST data was of low signal-to-noise, thus implying that only the most
severe blends were detected. In addition, the relatively small
sample used by \citetalias{A2003009} cannot be considered representative
of the Cepheid population in M\,31.

It is interesting to note that the results of \citetalias{A2007011},
based on a relatively large sample of Cepheids in M\,33, have mean values
similar to ours (although slightly different median values; Table
\ref{blends}). Such good agreement should be taken with caution as the
host galaxies, the methods used, the period distributions of the samples, and
therefore the associated systematic errors are different. Nevertheless it is
encouraging that estimates made using different methods in two somewhat
differing spiral galaxies yield similar results.

To further analyse the effects of blending we have applied several cuts to the
FM sample. It has traditionally been argued that the longer period (and
brighter) Cepheids should be less affected by blending \citep{A2006023}. We
computed the mean blending for Cepheids with a period longer than 12 days and
obtained, as expected, lower blending values (Table \ref{blends}). Furthermore,
considering that blending decreases the Cepheid amplitude, larger amplitude
Cepheids should also be less affected by blending and, in fact, the 37 Cepheids
with $\mathcal{A}_V>0.8$ mag do have lower blending values (Table
\ref{blends}).

Finally, a blending-corrected color excess can be computed from Eq.
(\ref{blendexcess}), obtaining a mean value of $<E(B^i-V^i)>= 0.305\pm0.011$.
The blended mean color excess is slightly lower ($<E(B-V)>=0.296\pm0.012$),
indicating that blending sources are bluer than Cepheids. The difference is
more significant when considering only Cepheids with large blending values
($\Delta_V<-0.5$ mag and $\Delta_B<-0.5$ mag), obtaining a blending-corrected
color excess of $<E(B^i-V^i)>= 0.472\pm0.028$, whereas
$<E(B-V)>=0.315\pm0.030$. Therefore, considering the color-magnitude diagram
(Fig. \ref{vbv}) and the distribution of Cepheids along the spiral arms, it is
likely that early-type and young main sequence stars are responsible for the
large measured blending. The obtained results are in good agreement
with the results in \citetalias{A2007011}, who also found that M\,33 blending
sources were on the average bluer than the Cepheids. To further check this
scenario we used young MW open clusters to compute the light contribution that
the main sequence stars would introduce on a Cepheid variable in the cluster.
The resulting predictions on $\Delta_V$ and $\Delta_B-\Delta_V$ are in good
agreement with the inferred blending values in M\,31.

\section{Distance determination}
\label{secdist}

As previously mentioned, long period and large amplitude Cepheids are less
affected by blends. Therefore, when computing the mean distance modulus to
M\,31, a systematic trend should be observed for increasing period and
amplitude cuts (i.e., rejecting short period or low amplitude Cepheids). The
$V_0-M_V$ values obtained after the metallicity correction (Sect.
\ref{secmetal}) were used to compute weighted mean distance determinations to
M\,31 (with $2.5\sigma$ clipping). The resulting values assume a distance
modulus to the LMC of $(m-M)_0=18.42\pm0.06$ mag (Sect. \ref{secabs}). The
amplitudes for the 41 FM Cepheids without accurate fits were computed from the
observations and also included in order to obtain a better coverage in periods
and amplitudes.

\begin{figure}[!tb]
 \resizebox{\hsize}{!}{\includegraphics*{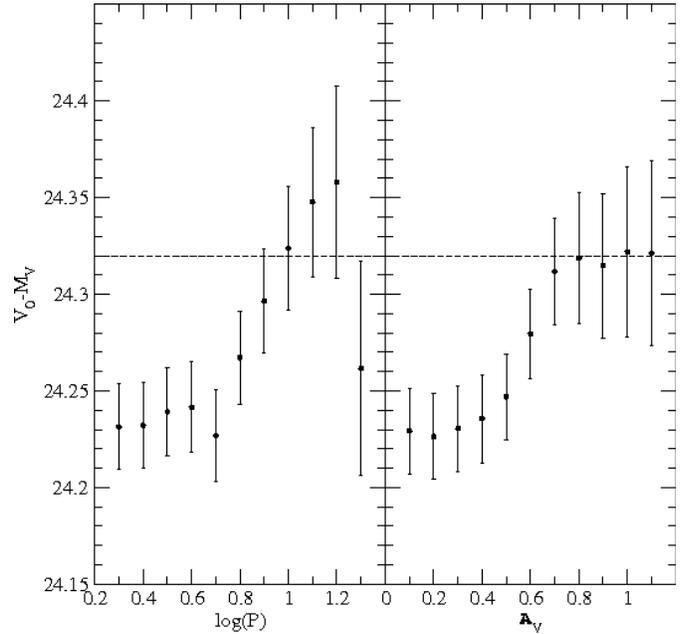}}
 \caption{Weighted mean $V_0-M_V$ values for the 281 fundamental mode 
Cepheids in M\,31 for different cuts. Error bars indicate the error on the 
mean. The dashed line represents the adopted distance determination of 
$(m-M)_0=24.32\pm0.12$ mag. Left: Each distance determination includes all 
Cepheids with period longer than the specified value. Right: Each distance 
determination includes all Cepheids with amplitude larger than the 
specified value.}
 \label{distance}
\end{figure}

Figure \ref{distance} shows that the distance modulus increases as the minimum
period or minimum amplitude cuts increase. We observe that the distance modulus
value stabilizes for $\mathcal{A}_V>0.8$ mag. The most likely cause for the
observed tendency is that blending is lower than photometric errors (or even
zero) for large amplitudes. The most suitable period cut is more difficult to
compute, although an increasing tendency is also observed. The observed
behavior is explained if long period Cepheids are still affected by large
blends, thus still introducing a bias in the derived distance modulus. Hence,
an amplitude cut is the best choice to derive the distance to M\,31 from the
studied Cepheid sample. From this analysis (Fig. \ref{distance}), the 66
Cepheids with amplitude $\mathcal{A}_V>0.8$ mag seem to represent the best
sample for distance determination.

\begin{figure}[!tb]
 \resizebox{\hsize}{!}{\includegraphics*{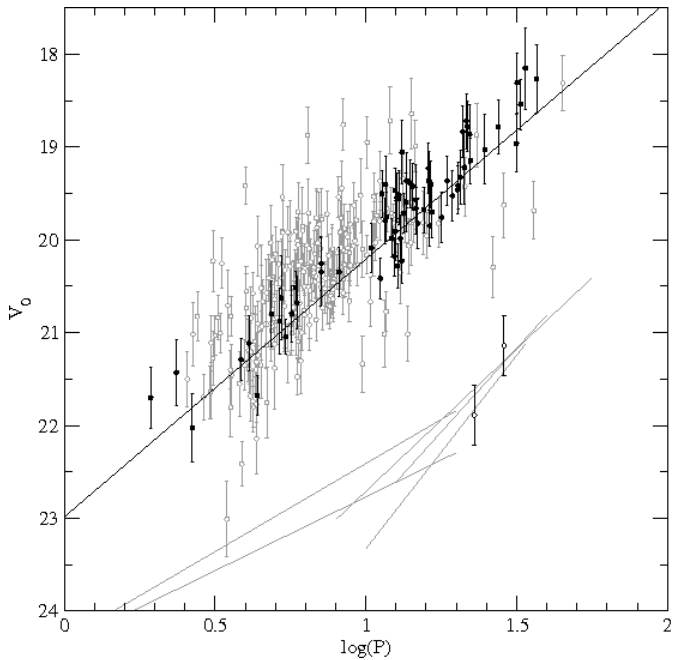}}
 \caption{Absorption-corrected $V$ magnitude as a function of period for 281
fundamental mode Cepheids. Empty grey circles: Complete fundamental
mode sample.  Filled black circles: Cepheids with $\mathcal{A}_V>0.8$ mag.
Empty black circles: Two stars with $\mathcal{A}_V>0.8$ mag excluded from the
distance determination. Black line: \citet{A2007006} P-L relationship, for a
distance modulus to M\,31 of 24.44 \citep{P2005002} and a mean metallicity
correction of $-0.1$ mag (Sect. \ref{secmetal}). Grey lines: P-L relationships
for type II Cepheids as given by \citet{Alcock98} at distance of M\,31.}
 \label{pvoamp}
\end{figure}

The effect of removing small amplitude Cepheids is clearly visible in Fig.
\ref{pvoamp}, where the P-L diagram for the largest amplitude Cepheids is
shown. Two stars with $\mathcal{A}_V>0.8$ mag are placed far below the general
P-L distribution (empty circles in Fig. \ref{pvoamp}). Both stars are on the
type II Cepheids relationships, reinforcing the hypothesis that these stars
are, in fact, type II Cepheids and they have not been considered from now on.
Finally, the obtained P-L slope, with a value of $-2.83\pm0.12$ mag~dex$^{-1}$,
is closer to the \citet{A2007006} LMC value of $-2.779\pm0.031$ mag~dex$^{-1}$
than to MW P-L slopes \citep[e.g.: $-3.087\pm0.085$
mag~dex$^{-1}$,][]{A2005011}, therefore favoring the adopted P-L relationships.

From the considerations above, the M\,31 distance modulus obtained from the
studied sample of Cepheids is
$(m-M)_0\simeq(V_0-M_V)_{\mathcal{A}>0.8}=24.32\pm0.12$ mag. This value is
compatible with most distance determinations found in the literature and the
assumed distance modulus in Sect. \ref{secblend}. Considering that some large
amplitude Cepheids may still be affected by blends, the distance modulus we
derive could have a slight negative bias. Note that our final value is 0.09 mag
larger than the weighted mean distance modulus for the 281 FM Cepheids
($(V_0-M_V)=24.23\pm0.12$ mag). Therefore, blending is clearly an important
effect that has to be considered when obtaining extragalactic distance
determinations. In fact, blending has nearly the same impact on the final
distance determination for M\,31 as the metallicity correction.

\section{Summary}
\label{seccon}

As a result of a variability survey in the North-Eastern quadrant of M\,31
\citepalias{P2006002}, 416 Cepheids were detected and measured with a large
number of epochs ($\sim 250$) per filter ($B$ and $V$). The large number of
detected Cepheids has allowed the direct study of their period distribution.
The results reveal that the period distributions in M\,31 and the MW are
extremely similar and, in addition, that the Cepheid sample obtained is almost
as complete as the DDO MW sample. The only difference may be the lack of long
period Cepheids.

The large number of epochs obtained in both filters permits an accurate
pulsation mode identification for 240 FM and 75 FO Cepheids. Although some FO
Cepheids were previously detected in M\,31 \citep{A2006021}, our sample
represents an important increase in the number of detected FO pulsators and
opens a new window to study the basic properties of these stars in another
Local Group galaxy. The sample of FM Cepheids was completed with 41 long period
(i.e., longer than 8 days) Cepheid variables, resulting in 281 FM Cepheids. The
posterior analysis revealed that at least three of these stars are type II
Cepheids.

The analysis of the P-L relationship for the FM Cepheids reveals a large
scatter, which is not explained solely through the effects of interstellar
absorption and metallicity. Although additional efforts are needed to reduce
the obtained uncertainties, a new method to compute the effect of blending is
presented. The exact dependence of the amplitudes among different passbands, as
well as more precise amplitude determinations, would greatly improve the
results shown here. Even by considering the large uncertainties, the large
number of studied Cepheids provides an accurate characterization of the mean
blending values. We conclude that the most likely cause of blending seems to be
the light from unresolved stars belonging to the same stellar associations or
clusters as Cepheid variables.

The effect of blending has been shown to be larger than 0.09 mag in the
distance modulus to M\,31, thus having an effect as important as the
metallicity correction. Therefore, blending should always be taken into account
when obtaining extragalactic distance determinations with Cepheids. It has also
been shown that an amplitude cut is more adequate to obtaining unbiased
distance determinations than a period cut, since the effect of blending is
better accounted for. Finally, the large amplitude Cepheids have been used to
obtain a distance modulus to M\,31 of $(m-M)_0=24.32\pm0.12$ mag.

\begin{acknowledgements}
The authors are very grateful to L.~M. Macri for useful comments during the
preliminary stage of this work. Thanks are also due to the anonymous
referee for providing valuable comments and information on additional blending
determinations. The David Dunlap Observatory and the OGLE teams are also
acknowledged for making all their data publicly available in FTP form. This
program was supported by the Spanish MCyT grant AyA2006-15623-C02-01/02. F.~V.
acknowledges support from the Universitat de Barcelona through a BRD
fellowship. I.~R. acknowledges support from the Spanish MEC through a Ram\'on
y Cajal fellowship.
\end{acknowledgements}

\bibliographystyle{aa}
\bibliography{bibliography}
\end{document}